\documentclass[aps,prb,reprint,superscriptaddress]{revtex4-2}
\usepackage{amsfonts, amssymb, amsmath,graphicx}
\usepackage{subfigure}
\usepackage{multirow}
\usepackage{dcolumn}
\usepackage{bm}
\usepackage{color}
\usepackage[colorlinks=true,allcolors=blue]{hyperref}

\begin{document}

\title{Relations between topology and the quantum metric for Chern insulators}

\author{Tomoki Ozawa}
\affiliation{Advanced Institute for Materials Research, Tohoku University, Sendai 980-8577, Japan}
\author{Bruno Mera}
\affiliation{Instituto de Telecomunica\c{c}\~oes, 1049-001 Lisboa, Portugal}
\affiliation{Departmento de F\'{i}sica, Instituto Superior T\'ecnico, Universidade de Lisboa, Av. Rovisco Pais, 1049-001 Lisboa, Portugal}
\affiliation{Departmento de Matem\'{a}tica, Instituto Superior T\'ecnico, Universidade de Lisboa, Av. Rovisco Pais, 1049-001 Lisboa, Portugal}

\date{\today}

\newcommand{\tom}[1]{{\color{red} #1}}
\definecolor{bbblue}{rgb}{0.,0.24,0.51}
\newcommand{\blue}{\color{bbblue}}

\begin{abstract}
We investigate relations between topology and the quantum metric of two-dimensional Chern insulators. The quantum metric is the Riemannian metric defined on a parameter space induced from quantum states. Similar to the Berry curvature, the quantum metric provides a geometrical structure associated to quantum states. We consider the volume of the parameter space measured with the quantum metric, which we call the quantum volume of the parameter space. We establish an inequality between the quantum volume of the Brillouin zone and that of the twist-angle space. Exploiting this inequality and the inequality between the Chern number and the quantum volume, we investigate how the quantum volume can be used as a good measure to infer the Chern number. The inequalities are found to be saturated for fermions filling Landau levels. Through various concrete models, we elucidate conditions when the quantum volume gives a good estimate of the topology of the system.
\end{abstract}

\maketitle

\section{Introduction}
Topological band structure and associated geometrical properties of quantum and classical states have become an important subject of study in a wide variety of disciplines in physics, ranging from solid-state electrons~\cite{Hasan:2010,Qi:2011,Chiu:2016}, where the concept of topological insulator was first invented~\cite{Klitzing:1980,Thouless:1982,Kohmoto:1985,Kane:2005a,Kane:2005b,Bernevig:2006}, to atomic, molecular, and optical (AMO) physics~\cite{Cooper:2019,Ozawa:2019RMP}, acoustics~\cite{Yang:2015,Huber:2016,Ma:2019}, active matter~\cite{Yamauchi:2020,Shankar:2020}, and even in geophysics~\cite{Delplace:2017}.

Topological properties, such as Chern numbers and winding numbers, characterize the global nature of the states defined in a suitable parameter space, such as momentum space. On the other hand, geometry characterizes local properties at each point in the parameter space. Topology and geometry are closely related to each other; for example, the integral of the Berry curvature, a geometrical quantity, over a two-dimensional compact surface gives rise to the topological Chern number. Recently, there is an increasing interest in the study of a geometrical property called the quantum metric~\cite{Provost:1980}. The quantum metric provides a positive semi-definite metric on a parameter space, and is known to be relevant in various phenomena such as the width of the maximally localized Wannier function~\cite{Marzari:1997,Marzari:2012,mat:ryu:10}, responses in dissipative systems~\cite{Neupert:2013,Albert:2016,Kolodrubetz:2017,Ozawa:2018a}, response upon periodical modulations~\cite{Ozawa:2018b,Ozawa:2019prr,Klees:2021}, non-adiabatic responses~\cite{Bleu:2018}, orbital magnetism~\cite{Raoux:2015,Gao:2015,Piechon:2016,Combes:2016,Freimuth:2017}, superfluid density~\cite{Peotta:2015,Julku:2016,Liang:2017,Iskin:2018}, Lamb-shift-like energy shift in excitons~\cite{Srivastava:2015}, Landau-level spreading in flat band systems~\cite{Rhim:2020}, and in the construction of fractional Chern insulators~\cite{cla:lee:tho:qi:dev:15,lee:cla:tho:17}.
Very recently, experimental measurements of the quantum metric have been reported in various systems~\cite{Asteria:2019,Tan:2019,Yu:2020,Klees:2020,Gianfrate:2020,Liu:2020}.

The aim of this paper is to study relations between the topological properties of the system and the quantum metric, focusing on two-dimensional Chern insulators.
Chern insulators are characterized by topology, in particular, by the Chern numbers of families of quantum states defined on the first Brillouin zone, which is the compact momentum space obtained in the presence of translation symmetry with respect to a lattice in real space. When considering many-body systems, another relevant parameter space is the twist-angle space; one considers a periodic boundary condition with a phase twist, and the many-body wave function is parametrized by the phase~\cite{Niu:1985}. When considering non-interacting fermions filling a band, the geometrical properties of the twist-angle space are given by an average of the corresponding quantities in momentum space, and hence the topological Chern numbers calculated in both spaces are identical. Besides the Chern number, we can construct a non-topological global quantity characterizing the geometrical structure of the states from the quantum metric.
Since the quantum metric provides a Riemannian structure in the parameter space, we can consider the volume of the parameter space measured using this metric; we call it quantum volume of the parameter space.
Inspired by the known inequality between the Chern number and the quantum volume, we investigate how and when the quantum volume can act as a useful measure to infer the topology of the system.
Furthermore, we establish an inequality between the quantum volume of momentum space and the quantum volume of the twist-angle space. We show that in the limit of bands approaching Landau levels, we can obtain a very good estimate of the topology (Chern number) just by studying the quantum volume.
The relations between topology and the quantum volume have intricate mathematical structures as well, which are analyzed in detail in the accompanying paper from the view point of K\"{a}hler geometry~\cite{MeraOzawa:published}.

The structure of the paper is as follows. In Sec.~\ref{sec:defs}, we introduce the quantum metric, the Berry curvature, and the Chern number for a general parameter space. In Sec.~\ref{sec:spaces}, we introduce the momentum space and the twist-angle space and discuss general relations between the topology and the quantum volumes. In Sec.~\ref{sec:models}, we then discuss specific models to see how the general relations can be applied to understand the topological features of various cases. We first analyze Landau levels, which are the energy levels of a two-dimensional system in the presence of a uniform magnetic field. We then discuss two-band models, taking the Haldane model~\cite{Haldane:1988} and the two-band model introduced in Yang {\it et al}.~\cite{Yang:2012} as explicit examples. As we will see, two-band models have characteristic features which are not present in models with more than two bands. We finally discuss multi-band models, taking the Harper-Hofstadter model~\cite{Harper:1955,Hofstadter:1976} and the three-band models introduced in Ref.~\onlinecite{Sun:2011} and Ref.~\onlinecite{Yang:2012} as examples. We finally give conclusions in Sec.~\ref{sec:conclusion}. Detailed derivations of some formulas and relations are given in the Appendix.

\section{Definitions and Properties}
\label{sec:defs}
\subsection{Definitions}
We first define the quantum metric and the Berry curvature for a set of mutually orthogonal quantum states $|\psi_m (\boldsymbol\lambda)\rangle$, labeled by an integer $m = 1, 2, \cdots$, which depend on a set of parameters $\boldsymbol\lambda = (\lambda_1, \lambda_2, \cdots)$.
In what follows, $\boldsymbol\lambda$ will either be a quasimomentum $\mathbf{k}$ or the twist angle $(\theta_x, \theta_y, \cdots)$.

The geometrical structure of a set of quantum states $|\psi_m(\boldsymbol\lambda)\rangle$ with $m = 1, 2, \cdots, r$, where $r$ is the number of occupied bands for band insulators, is characterized by the quantum geometric tensor, whose real part gives the quantum metric and the imaginary part gives the Berry curvature~\cite{Provost:1980}. The quantum geometric tensor for a set of states $|\psi_m (\boldsymbol\lambda)\rangle$ with $1 \le m \le r$ is defined by~\cite{Kolodrubetz:2017}
\begin{align}
	\chi_{ij}(\boldsymbol\lambda)
	&\equiv
	\sum_{m = 1}^r
	\langle \partial_{\lambda_i} \psi_m (\boldsymbol\lambda) | (1 - P(\boldsymbol\lambda)) |\partial_{\lambda_j} \psi_m (\boldsymbol\lambda)\rangle,
\end{align}
where $P(\boldsymbol\lambda) \equiv \sum_{m = 1}^r |\psi_m (\boldsymbol\lambda) \rangle\langle \psi_m (\boldsymbol\lambda)|$ is the projector onto the set of states with $m = 1, 2, \cdots, r$.
The quantum metric $g_{ij}$ and the Berry curvature $\Omega_{ij}$ are defined through the real and imaginary parts of the quantum geometric tensor as
\begin{align}
	g_{ij}(\boldsymbol\lambda) &\equiv \mathrm{Re} [ \chi_{ij}(\boldsymbol\lambda) ].
	\\
	\Omega_{ij}(\boldsymbol\lambda) &\equiv - 2\, \mathrm{Im} [ \chi_{ij}(\boldsymbol\lambda) ].
\end{align}
It is straightforward to show $g_{ij} = g_{ji}$ and $\Omega_{ij} = -\Omega_{ji}$ from the definition of the quantum geometric tensor.

In this paper, we focus on two-dimensional systems: $\boldsymbol\lambda = (\lambda_1, \lambda_2)$. Then, in matrix notation,
\begin{align}
	\chi (\boldsymbol\lambda)
	=
	\begin{pmatrix}
	g_{11}(\boldsymbol\lambda) & g_{12}(\boldsymbol\lambda) - i\Omega_{12}(\boldsymbol\lambda)/2 \\
	g_{12}(\boldsymbol\lambda) + i\Omega_{12}(\boldsymbol\lambda)/2 & g_{22}(\boldsymbol\lambda)
	\end{pmatrix}.
\end{align}

\subsection{Basic properties}
The quantum metric $g_{ij}(\boldsymbol\lambda)$ provides a positive semidefinite metric in the parameter space described by $\boldsymbol\lambda = (\lambda_1, \lambda_2)$ satisfying $\det (g) = g_{11}g_{22} - g_{12}^2 \ge 0$. We note that the quantum metric is not a positive definite metric; there can be points in the parameter space where $\det (g) = 0$ holds.
The Berry curvature $\Omega_{ij}(\boldsymbol\lambda) d\lambda_i \wedge d\lambda_j$ provides the curvature $2$-form of a connection on a vector bundle whose fiber at $\boldsymbol\lambda$ is spanned by the set of states $|\psi_m (\boldsymbol\lambda)\rangle$ with $m = 1, 2, \cdots, r$.

Even though the quantum metric is a geometrical property of the tangent bundle of the parameter space and the Berry curvature is a property of the fiber bundle made of the quantum states, there are relations between the two; after all, they are both defined in terms of the same family of quantum states. An important relation that we want to explore in detail in this paper is the inequality that holds between the volume form and the Berry curvature:
\begin{align}
	\sqrt{\det (g)} \ge \frac{|\Omega_{12}|}{2}. \label{eq:ineq}
\end{align}
This inequality is well known in the literature and is due to the fact that the quantum geometric tensor $\chi_{ij}$ is a positive semi-definite matrix; the inequality was first shown for the single-band case ($r=1$) in Roy~\cite{Roy:2014}, and then later generalized to include multi-band cases by Peotta and T{\"o}rm{\"a}~\cite{Peotta:2015}.
For completeness, we give a proof of this inequality in Appendix~\ref{sec:app1} using Cauchy-Schwarz inequality, clarifying the condition under which the equality holds.
By integrating this inequality, we obtain an important relation between the topological Chern number and the quantum volume as we describe in the following.
The Chern number can be calculated by integrating the Berry curvature over the parameter space (when the parameter space is compact, which will be the case below):
\begin{align}
	\mathcal{C}
	=
	\frac{1}{2\pi}\int d\lambda_1 d\lambda_2 \; \Omega_{12}(\boldsymbol\lambda).
\end{align}
The Chern number is guaranteed to be an integer and reflects a topological property of the states.
The quantum volume $\mathrm{vol}_g$ is the volume of the parameter space calculated using the quantum metric as the metric, and it is defined by
\begin{align}
    \mathrm{vol}_g &\equiv \int d\lambda_1 d\lambda_2 \sqrt{\det (g)}.
\end{align}
Using the inequality Eq.~(\ref{eq:ineq}), the following inequality between the quantum volume and the Chern number holds~\cite{Roy:2014,Peotta:2015}:
\begin{align}
	\mathrm{vol}_g
	\ge \int d\lambda_1 d\lambda_2 \frac{|\Omega_{12}|}{2}
	\ge \left| \int d\lambda_1 d\lambda_2 \frac{\Omega_{12}}{2}\right|
	=
	\pi |\mathcal{C}|. \label{eq:metricchern}
\end{align}
Thus, the Chern number multiplied by $\pi$ is upper bounded by the quantum volume of the parameter space.
This relation has, for example, been applied to topological superconductors to find a relation between the superfluid weight and the topological properties of the system~\cite{Peotta:2015}.
In this paper, we will investigate consequences of this inequality applied to the quasimomentum and twist-angle spaces as relevant parameter spaces.
Before studying concrete parameter spaces, we briefly discuss how the quantum geometric tensor is related to the Hamiltonian of the system.

\subsection{Relation to the Hamiltonian}
So far we have not introduced any Hamiltonian. Typically, quantum states $|\psi_m (\boldsymbol\lambda)\rangle$ are defined as eigenstates of a certain Hamiltonian $H(\boldsymbol\lambda)$, which depends smoothly on the same parameter $\boldsymbol\lambda$. We then define that $|\psi_m (\boldsymbol\lambda)\rangle$ is an $m$th eigenstate of $H(\boldsymbol\lambda)$, where the corresponding eigenvalue $\epsilon_m (\boldsymbol\lambda)$ is the $m$th smallest of all the eigenvalues of $H(\boldsymbol\lambda)$. The eigenvalue equation is $H(\boldsymbol\lambda) |\psi_m (\boldsymbol\lambda)\rangle = \epsilon_m (\boldsymbol\lambda) |\psi_m (\boldsymbol\lambda)\rangle$. Using the Hamiltonian, we can find an expression for the quantum geometric tensor, which does not involve the derivative of quantum states. From the definition of the quantum geometric tensor, we see
\begin{align}
	&\chi_{ij}(\boldsymbol\lambda)
	\notag \\
	&=
	\sum_{m = 1}^r\sum_{m^\prime> r}
	\langle \partial_{\lambda_1}\psi_m (\boldsymbol\lambda) | \psi_{m^\prime} (\boldsymbol\lambda)\rangle \langle \psi_{m^\prime} (\boldsymbol\lambda)| \partial_{\lambda_j} \psi_m (\boldsymbol\lambda) \rangle.
\end{align}
Using the relation $\langle \psi_{m^\prime} (\boldsymbol\lambda) | \frac{\partial H}{\partial \lambda_i} | \psi_m (\boldsymbol\lambda)\rangle = (\epsilon_m (\boldsymbol\lambda) - \epsilon_{m^\prime} (\boldsymbol\lambda)) \langle \psi_{m^\prime} (\boldsymbol\lambda) | \partial_{\lambda_i} \psi_m (\boldsymbol\lambda) \rangle$, we can rewrite the quantum geometric tensor as
\begin{align}
	&\chi_{ij}(\boldsymbol\lambda)
	\notag \\
	&=
	\sum_{m=1}^r
	\sum_{m^\prime > r} \frac{\langle \psi_m (\boldsymbol\lambda) | \frac{\partial H}{\partial \lambda_i} | \psi_{m^\prime} (\boldsymbol\lambda)\rangle \langle \psi_{m^\prime} (\boldsymbol\lambda) | \frac{\partial H}{\partial \lambda_j} | \psi_m (\boldsymbol\lambda)\rangle}{(\epsilon_{m^\prime} (\boldsymbol\lambda) - \epsilon_m (\boldsymbol\lambda))^2}. \label{eq:usefulformula}
\end{align}
This is an expression which does not involve any derivatives of the quantum states themselves, and is useful when numerically calculating the quantum geometric tensor.
We note that there is also a technique of calculating the quantum geometric tensor without computing the eigenstates at all as discussed in Refs.~\cite{Pozo:2020,Graf:2021}.

\section{Momentum space and twist-angle space}
\label{sec:spaces}
Now we explicitly consider momentum space and twist-angle space as parameter spaces which are relevant in the study of two-dimensional Chern insulators.
An $n$-band model is characterized by an $n$-by-$n$ Hamiltonian $H(k_x, k_x)$, which depends on quasi momenta $(k_x, k_y)$ defined in the first Brillouin zone, which is periodic and is, topologically, a two-torus. The quantum state $|\psi_m (k_x,k_y)\rangle$, often called the Bloch state, is the eigenvector of $H(k_x,k_y)$ with the $m$-th smallest eigenvalue.
From the Bloch states with $r$ smallest eigenvalues, we construct the quantum metric and the Berry curvature, which we write as $g_{ij}(k_x,k_y)$ and $\Omega_{ij}(k_x,k_y)$.
Topology and geometry of the Bloch states are directly relevant in the study of topological insulators; Chern insulators are two-dimensional systems where the Chern number of the Bloch states, associated with the filled bands, over momentum space is nonzero. An important physical consequence of nonzero Chern number is the bulk-edge correspondence; the number of edge modes of a given energy gap is equal to the sum of the Chern numbers of energy bands below the gap~\cite{Hatsugai:1993a,Hatsugai:1993b}.

Another natural parameter space to analyze Chern insulators is the twist-angle space~\cite{Niu:1985}, which is more abstract than momentum space but can be applied also to disordered and interacting many-body systems.
The geometrical structure in the twist-angle space is defined by introducing a periodic boundary condition with a twist phase $\theta_x$ and $\theta_y$ along $x$ and $y$ directions, respectively, and defining quantum states in the parameter space of $(\theta_x, \theta_y)$.
Twisted boundary conditions can be gauged away by a suitable gauge transformation, but at a cost of changing the Hamiltonian from $H(k_x,k_y)$ to $H(k_x + \theta_x/L_x, k_y + \theta_y/L_y)$, where $L_x$ and $L_y$ are the lengths of the system in $x$ and $y$ directions, respectively. It then follows that $\frac{\partial H}{\partial k_i} = L_i \frac{\partial H}{\partial \theta_i}$.
Then, from the expression of the quantum geometric tensor in terms of the derivative of the Hamiltonian, Eq.~(\ref{eq:usefulformula}), we can see that the quantum geometric tensor in twist-angle space is $1/L_iL_j$ times the quantum geometric tensor defined in momentum space.
We are most interested in the situation where $r$ bands labeled by $m = 1, 2, \cdots, r$ are completely filled. We let $\tilde{\chi}_{ij}(\theta_x, \theta_y)$ denote the quantum geometric tensor in twist-angle space when $r$ bands are filled. Assuming non-interacting fermions, the quantum geometric tensor for the entire band is a sum of the quantum geometric tensor of each quasimomentum, namely
\begin{align}
	\tilde{\chi}_{ij}(\theta_x, \theta_y) & = \frac{1}{L_i L_j}\sum_\mathbf{k} \chi_{ij}(k_x, k_y)
	\notag \\
	&=\frac{1}{(2\pi)^2}\int d^2k \chi_{ij}(k_x, k_y),
\end{align}
which shows that the quantum geometric tensor for a set of bands in the twist-angle space is nothing but the average of the quantum geometric tensor defined in momentum space. One can show that this integral is $\theta_x$ and $\theta_y$ independent in the thermodynamic limit~\cite{Watanabe:2018,Kudo:2019}.
Explicitly writing out each component of the quantum geometric tensor, we obtain
\begin{align}
	\tilde{g}_{ij}(\theta_x, \theta_y) &= \frac{1}{(2\pi)^2}\int d^2 k\; g_{ij}(k_x, k_y), \\
	\tilde{\Omega}_{xy}(\theta_x, \theta_y) &= \frac{1}{(2\pi)^2}\int d^2 k\; \Omega_{xy}(k_x, k_y) = \frac{1}{2\pi} \mathcal{C}, \label{eq:ktheta}
\end{align}
where $\tilde{g}_{ij}$ and $\tilde{\Omega}_{ij}$ are the quantum metric and the Berry curvature for the filled $r$ bands defined in the twist-angle space. We note again that these quantities do not depend on the twist angles in the thermodynamic limit, or in other words, these quantities are flat in the twist-angle space. As it is clear from the relation Eq.~(\ref{eq:ktheta}), the Chern number calculated from $\Omega_{xy}(k_x, k_y)$ and $\tilde{\Omega}_{xy}(\theta_x, \theta_y)$ are the same:
\begin{align}
	\mathcal{C} = \frac{1}{2\pi}\int d^2 k\; \Omega_{xy}(k_x,k_y) = \frac{1}{2\pi}\int d^2 \theta \; \tilde{\Omega}_{xy}(\theta_x, \theta_y).
\end{align}
Therefore, for non-interacting fermions completely filling a set of bands, the Chern number in twist-angle space gives the same information as the Chern number in momentum space.
On the other hand, the quantum volumes of momentum space and the twist-angle space are generally different.
We note that the quantum metric in twist-angle space is proportional to the localization tensor of the material~\cite{sou:wil:mar:00}, which is directly related to the longitudinal response of the system upon periodic modulation when a band is filled~\cite{Asteria:2019,Ozawa:2019prr}. The quantum volume of twist-angle space has the interpretation of the geometric mean of the eigenvalues of the localization tensor, giving an average notion of the spread of the many-particle ground state in real space~\cite{mer:20:1}.

Since the Chern numbers for non-interacting fermions are the same for momentum space and the twist-angle space, the inequality Eq.(\ref{eq:metricchern}) holds for both spaces with the same Chern number:
\begin{align}
	\pi |\mathcal{C}| &\le \mathrm{vol}_g \equiv \int dk_x dk_y \sqrt{\mathrm{det}[g(k_x, k_y)]}, \notag \\
	\pi |\mathcal{C}| &\le \mathrm{vol}_{\tilde{g}} \equiv \int d\theta_x d\theta_y \sqrt{\mathrm{det}[]\tilde{g}(\theta_x, \theta_y)]}.
\end{align}
The quantum volumes $\mathrm{vol}_g$ and $\mathrm{vol}_{\tilde{g}}$ are generally different, and in fact the inequality $\mathrm{vol}_g \le \mathrm{vol}_{\tilde{g}}$ holds; the quantum volume of momentum space is always smaller or equal to the quantum volume of the twist-angle space. The proof of $\mathrm{vol}_g \le \mathrm{vol}_{\tilde{g}}$ is given in Appendix~\ref{sec:gg}.
Combining all the inequalities, we finally obtain
\begin{align}
	\pi |\mathcal{C}| \le \mathrm{vol}_g \le \mathrm{vol}_{\tilde{g}}, \label{eq:main}
\end{align}
which is the central relation between the topology and the quantum metric whose consequence we will explore in the paper.
In the accompanying paper~\cite{MeraOzawa:published}, this inequality is analyzed from the mathematical viewpoint of K\"{a}hler geometry. In particular, it is discussed that the first inequality is saturated, provided the quantum metric on momentum-space is non-degenerate, when the parameter space equipped with the quantum metric and the Berry curvature, playing the role of a symplectic form, forms a K\"{a}hler manifold.

One immediate consequence of this inequality is that if $\mathrm{vol}_g < \pi $ or $\mathrm{vol}_{\tilde{g}} < \pi$, then $|\mathcal{C}| = 0$, i.e., a set of bands is topologically trivial.
For systems with $\mathrm{vol}_g \ge \pi $, we cannot make a definite statement about their topology. However, as we confirm below with several models, $\mathrm{vol}_g / \pi $ often gives a very good estimate of the topology, and we can also see that in a proper limit approaching the situation similar to the filled Landau levels, $\mathrm{vol}_g / \pi $ approaches the Chern number from above.

\section{Models}
\label{sec:models}

Now we explore how the topology and geometry are related through various examples and models.
We first consider the paradigmatic case of Landau levels associated to the problem of two-dimensional electrons in the presence of a uniform magnetic field. We then consider two-band models, and then finally discuss multi-band models.

\subsection{Landau levels}
A charged particle in a uniform magnetic field is the simplest example of a system with nontrivial Chern number.
The Hamiltonian of a charged particle with mass $M$ in a uniform magnetic fields takes the following form:
\begin{align}
	H = \frac{\hat{p}_x^2 + (\hat{p}_y - B\hat{x})^2}{2M},
\end{align}
where we assumed that the charge is unity, and we are taking the Landau gauge for the magnetic vector potential $\vec{A}(\mathbf{r}) = (0, Bx)$.
The Hamiltonian does not have a translational symmetry, but there is a magnetic translational symmetry, which is a symmetry of a discrete translation followed by a phase twist. To understand that there is a magnetic translational symmetry, we consider a fictitious magnetic unit cell, which is a rectangle with lengths $a_x$ and $a_y$ in $x$ and $y$ directions, respectively, which satisfies $|B| a_x a_y = 2\pi$. Note that the choice of $a_x$ and $a_y$ is not unique. Using the magnetic translational symmetry, we can apply Bloch's theorem and introduce quasimomentum, which takes a value in the Brillouin zone: $-\pi/a_x \le k_x \le \pi/a_x$ and $-\pi/a_y \le k_y \le \pi/a_y$.

The quantum metric and the Berry curvature of each Landau level have been evaluated in Ref.~\cite{Peotta:2015} by taking an appropriate limit of the Harper-Hofstadter model.
We give an alternative derivation of the quantum geometric tensor of the Landau levels in Appendix~\ref{sec:ll} directly using the wave functions of the Landau levels, which allow us to derive formulas for the quantum geometric tensor of multiple Landau levels.
The quantum metric and the Berry curvature for the collection of lowest $r$ Landau levels is
\begin{align}
	g (k_x, k_y) &= \frac{r}{2|B|} \begin{pmatrix} 1 & 0 \\ 0 & 1\end{pmatrix},
	&
	\Omega (k_x, k_y) &= -\frac{r}{B}.
\end{align}
Thus, the quantum geometric tensor is completely flat in the Brillouin zone.
This flatness implies that the quantum geometric tensor $\chi_{ij}$ of the Brillouin zone and $\tilde{\chi}_{ij}$ of the twist-angle space coincide.
The Chern number and the quantum volumes are $\mathcal{C} = -r\ \mathrm{sign}(B)$, $\mathrm{vol}_g = \mathrm{vol}_{\tilde{g}} =  r \pi$ and thus, the equality
\begin{align}
	\pi |\mathcal{C}| = r \pi = \mathrm{vol}_g = \mathrm{vol}_{\tilde{g}}
\end{align}
holds for the collection of $r$ lowest Landau levels.
As we see below for the case of the multi-band models, we can heuristically say that, as the system approaches the Landau levels, the equality is approached.

\subsection{Two-band models}

There are two special properties that only hold for two-band models, where the quantum states are just $|\psi_1 (\boldsymbol\lambda)\rangle$ and $|\psi_2 (\boldsymbol\lambda)\rangle$, and we consider geometrical properties of one of the two states.
The first property is:
\begin{quote}
	When the parameter space is a two-torus, there must be some points in the parameter space where $\det (g) = |\Omega_{12}| = 0$.
\end{quote}
This property and its proof are presented in the accompanying paper~\cite{MeraOzawa:published} as Theorem~3, exploiting the difference in the fundamental groups of the two-torus and a complex projective space, which, in this case, is just a two-sphere.
A consequence of this property is that, for two-band Chern insulators with a nontrivial Chern number, one cannot achieve a situation where either the quantum metric or the Berry curvature becomes flat in the parameter space, which has also been noted in Ref.~\cite{lee:cla:tho:17}. This implies that it is not possible to achieve the limit of lowest Landau levels with two-band models.
On the other hand, there is a second property which holds only for two-band models, which is:
\begin{quote}
	The equality $\displaystyle \sqrt{\det(g)} = \frac{|\Omega_{12}|}{2}$ always holds.
\end{quote}
This relation has been noticed by several earlier works~\cite{Ma:2013, Liang:2017, Ma:2020}.
One can see this property by confirming that the condition for the saturation of the inequality Eq.~(\ref{eq:main}) always holds for two-band models as noticed in Ref.~\cite{Ma:2020} and also detailed in Appendix~\ref{sec:app1}, or also by calculating directly the volume form and the Berry curvature for two-band models, as in Refs.~\cite{Ma:2013, Liang:2017}.
A simple consequence of the second property is that, if $\Omega_{12}$ is everywhere non-negative in the parameter space, $\mathrm{vol}_g = \pi \mathcal{C}$, and if $\Omega_{12}$ is everywhere nonpositive in the parameter space, $\mathrm{vol}_g = -\pi \mathcal{C}$. This implies that the quantum volume directly tells us about the topology of the quantum states if there is no change of sign of the Berry curvature over the parameter space. We will see shortly that it is often the case that the Berry curvature does not change its sign in the Brillouin zone, and thus the quantum volume coincides with $\pi |\mathcal{C}|$.

\subsubsection{Haldane model}

\begin{figure}[htbp]
\begin{center}
\includegraphics[width= 0.4 \textwidth]{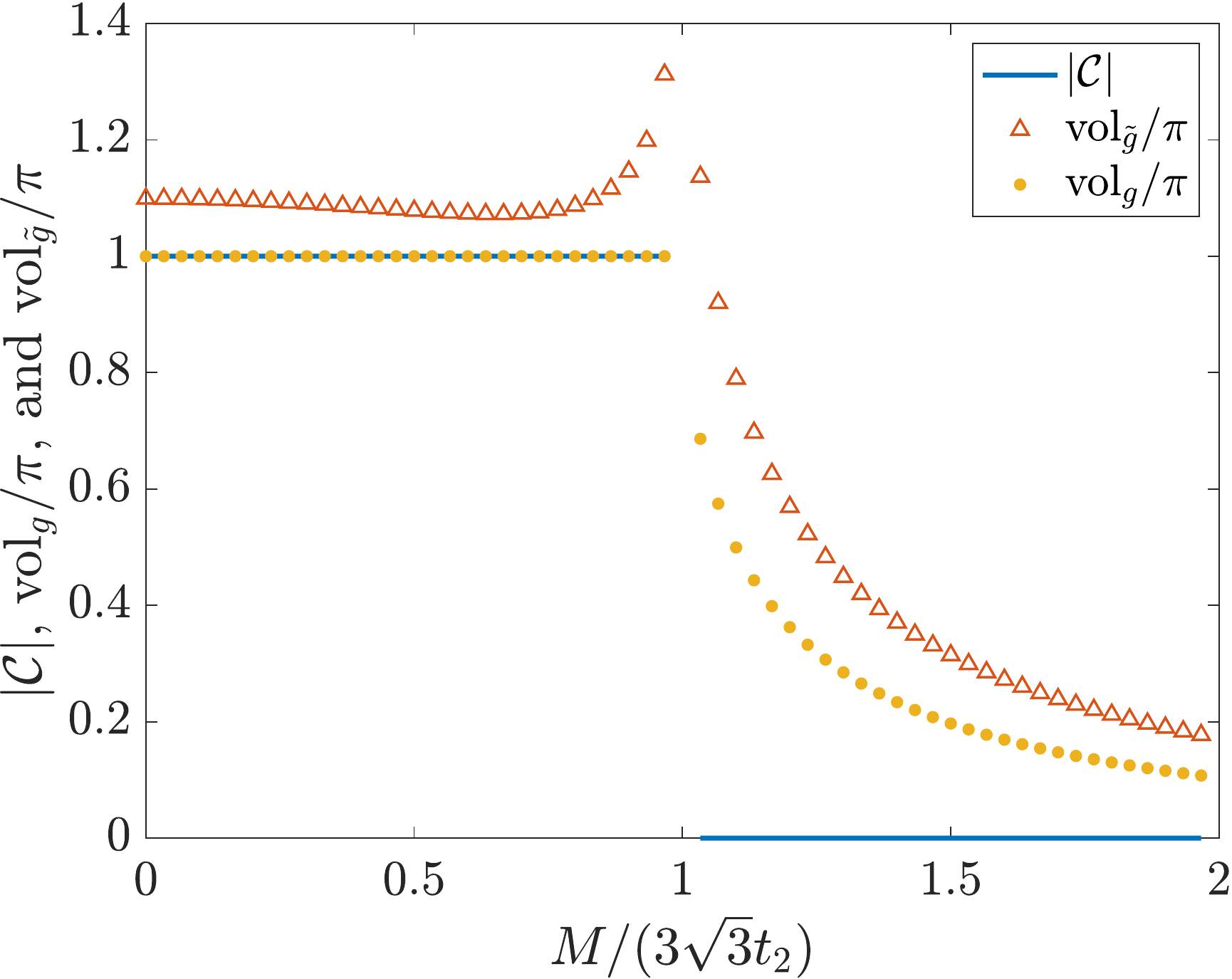}
\caption{The Haldane model: The Chern number and the quantum volumes as a function of a parameter $M$ in the model. The solid line corresponds to the Chern number, and the circles and triangles are $\mathrm{vol}_g/\pi$ and $\mathrm{vol}_{\tilde{g}}/\pi$, respectively. There is a topological phase transition at $M/(3\sqrt{3}t_2) = 1$.}
\label{fig:haldane}
\end{center}
\end{figure}

As an explicit example of a two-band model, we first analyze the Haldane model~\cite{Haldane:1988}.
The Haldane model is a tight-binding model on a honeycomb lattice with complex next-nearest-neighbor hopping amplitudes.
The Haldane model has various topological phases as one varies $\phi$ and $M/(3\sqrt{3}t_2)$, where $\phi$ is the phase of next-nearest-neighbor hopping, and $t_2$ is its magnitude. $M$ is the on-site energy difference between the two sublattices within a unit cell.
The Hamiltonian is
\begin{align}
	&H(\mathbf{k}) =
	\notag \\
	&\begin{pmatrix}
	2t_2 \sum_j \cos (\phi - \mathbf{k} \cdot \mathbf{R}_j^\prime) + M & t_1 \sum_j e^{-i\mathbf{k}\cdot \mathbf{R}_j} \\
	t_1 \sum_j e^{i\mathbf{k}\cdot \mathbf{R}_j} & 2t_2 \sum_j \cos (\phi + \mathbf{k}\cdot \mathbf{R}_j^\prime) - M
	\end{pmatrix},
\end{align}
where $t_1$ is the nearest-neighbor hopping, $\mathbf{R}_j$ and $\mathbf{R}_j^\prime$ with $j =$ 1, 2, and 3 are the vectors connecting nearest- and next-nearest-neighbor sites in the honeycomb lattice.
In Fig.~\ref{fig:haldane}, we plot $|\mathcal{C}|$, $\mathrm{vol}_g/\pi$, and $\mathrm{vol}_{\tilde{g}}/\pi$ as a function of $M/(3\sqrt{3}t_2)$ for a fixed value of $\phi = \pi/2$. There is a topological phase transition at $M/(3\sqrt{3}t_2) = 1$.

The figure shows that, in the topologically nontrivial regime, $|\mathcal{C}| = \mathrm{vol}_g/\pi = 1$ holds, which is a consequence of a fact that $\sqrt{\det (g)} = |\Omega_{12}|/2$ strictly holds in the case of two-band models, together with the fact that the Berry curvature does not change sign within the Brillouin zone for the Haldane model.
In the topologically trivial regime, we see $\mathrm{vol}_g/\pi < 1$, which implies the correct relation $|\mathcal{C}| = 0$. Therefore, the quantum volume in momentum space precisely captures the topology of the model.

We also note that the quantum volume of the twist-angle space, $\mathrm{vol}_{\tilde{g}}/\pi$, is not equal to the Chern number in the topological regime. This is because the relation $\sqrt{\det (g)} = |\Omega_{12}|/2$ holds if the Hilbert space of quantum states for a particular value of a parameter is two-dimensional, but in the twist-angle space, the relevant Hilbert space is a many-body Fock space; the Hilbert space is much larger than two even for two-band models.

\subsubsection{Square lattice model with higher Chern numbers}

\begin{figure}[htbp]
\begin{center}
\includegraphics[width= 0.4 \textwidth]{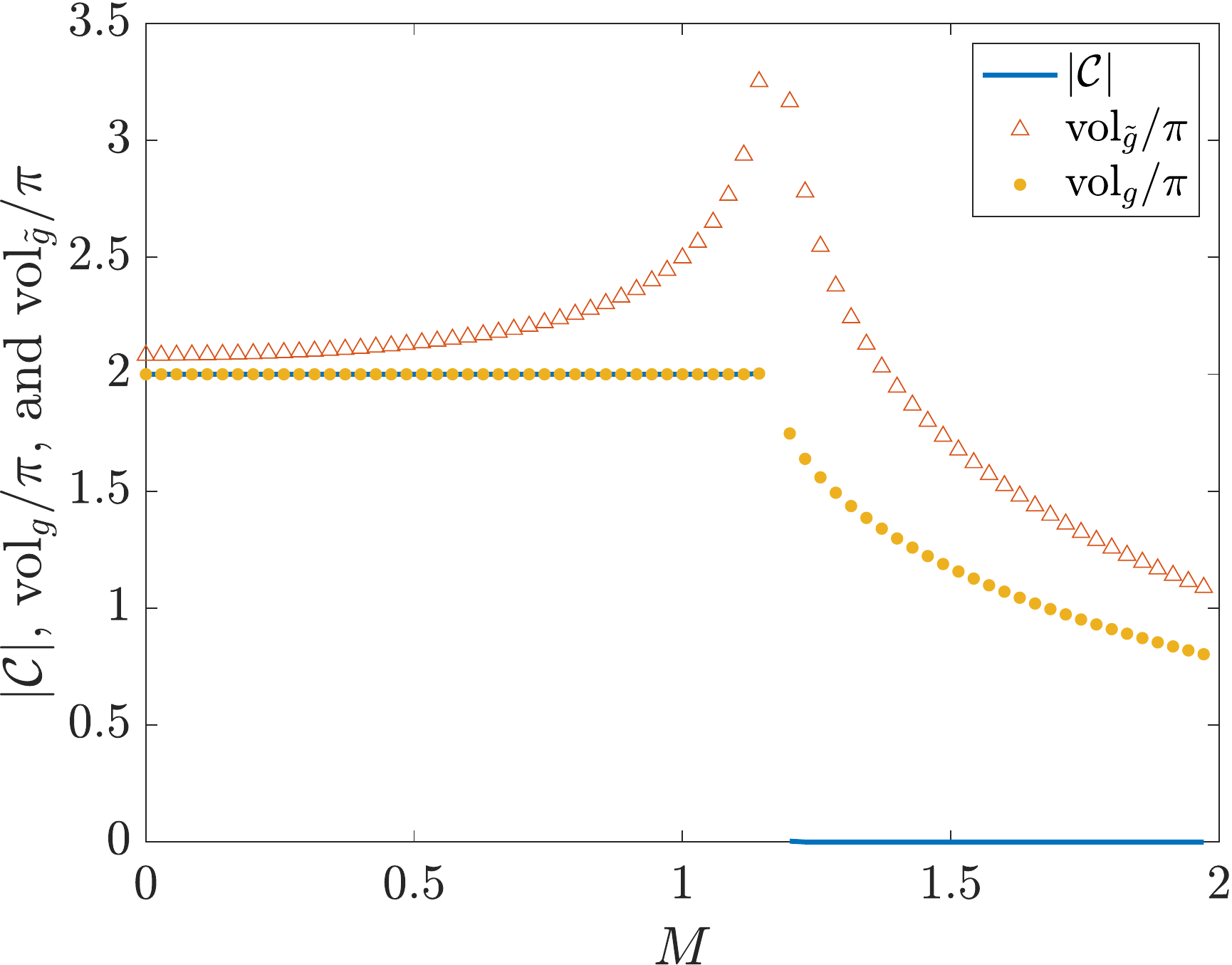}
\caption{The square-lattice two-band model with higher Chern numbers from Yang {\it et al}.~\cite{Yang:2012}: The Chern number and the quantum volumes as a function of a parameter $M$ in the model. The solid line corresponds to the Chern number, and the circles and triangles are $\mathrm{vol}_g/\pi$ and $\mathrm{vol}_{\tilde{g}}/\pi$, respectively. There is a topological phase transition around $M \approx 1.3$.}
\label{fig:yang2band}
\end{center}
\end{figure}

Next, as an example of a topological two-band model with the Chern number greater than one, we consider the square lattice model introduced in Yang {\it et al}.~\cite{Yang:2012}, which has two orbital degrees of freedom per site.
The Hamiltonian is
\begin{widetext}
\begin{align}
    H(\mathbf{k})=
    \begin{pmatrix}
    2 t_3 (\cos (2k_x) + \cos (2k_y)) - 4t_2 \sin (k_x) \sin (k_y)+M
    & 2 t_1 (e^{i\pi/4} \cos (k_x) + e^{-i\pi/4} \cos (k_y))
    \\
    2 t_1 (e^{-i\pi/4} \cos (k_x) + e^{i\pi/4} \cos (k_y))
    &
    2 t_3 (\cos (2k_x) + \cos (2k_y)) + 4t_2 \sin (k_x) \sin (k_y) - M
    \end{pmatrix}
\end{align}
\end{widetext}
The orbital-dependent energy $M$ is not in the original paper of Yang {\it et al}.~\cite{Yang:2012}; we add this factor to induce a topological phase transition.
For other parameters, we take $t_1 = 1$, $t_2 = 1/(2+\sqrt{2})$, and $t_3 = 1/(2+2\sqrt{2})$ as in Ref.~\cite{Yang:2012}.
When $M=0$, the lower (upper) band has the Chern number of $-2$($+2$). As one increases $M$, the bands go through a topological phase transition and become topologically trivial.
As a function of $M$, we plot the Chern number and the quantum volumes in Fig.~\ref{fig:yang2band}.
We see that the equality $|\mathcal{C}| = \mathrm{vol}_g/\pi = 2$ is again achieved in the topological region.

Although achieving $|\mathcal{C}| = \mathrm{vol}_g/\pi$ requires that the Berry curvature does not change its sign over the Brillouin zone, we have confirmed through these examples that this equality is often realized in topological two-band models.

\subsection{Multi-band models}
For multi-band systems, we do not have an equality $\sqrt{\det(g)} = |\Omega_{12}|/2$ in general. On the other hand, we do not have the constraint that $\sqrt{\det(g)}$ and $|\Omega_{12}|$ vanish somewhere in the Brillouin zone. The latter property implies that we can consider the limit where the quantum metric and the Berry curvature become flat (constant) over the Brillouin zone, which is similar to the situation of the Landau levels. For examples of multi-band models, we consider the Harper-Hofstadter model~\cite{Harper:1955,Hofstadter:1976}, the three-band model with a unit Chern number introduced in Sun {\it et al}.~\cite{Sun:2011}, and the three-band model with a higher Chern number introduced in Yang {\it et al}.~\cite{Yang:2012}.

\subsubsection{Harper-Hofstadter model}

The Harper-Hofstadter model is a square-lattice tight-binding model with a magnetic flux $\phi$.
When $\phi = 2\pi p/q$ with mutually prime integers $p$ and $q$, the model has $q$ bands.
We consider the lowest one and two bands of the Harper-Hofstadter model for $p = 1$ to show that, in the large $q$ limit, the model approaches the limit of the Landau levels and the inequality Eq.~(\ref{eq:main}) is saturated.

\begin{figure*}[htbp]
\begin{center}
\includegraphics[width= 0.4 \textwidth]{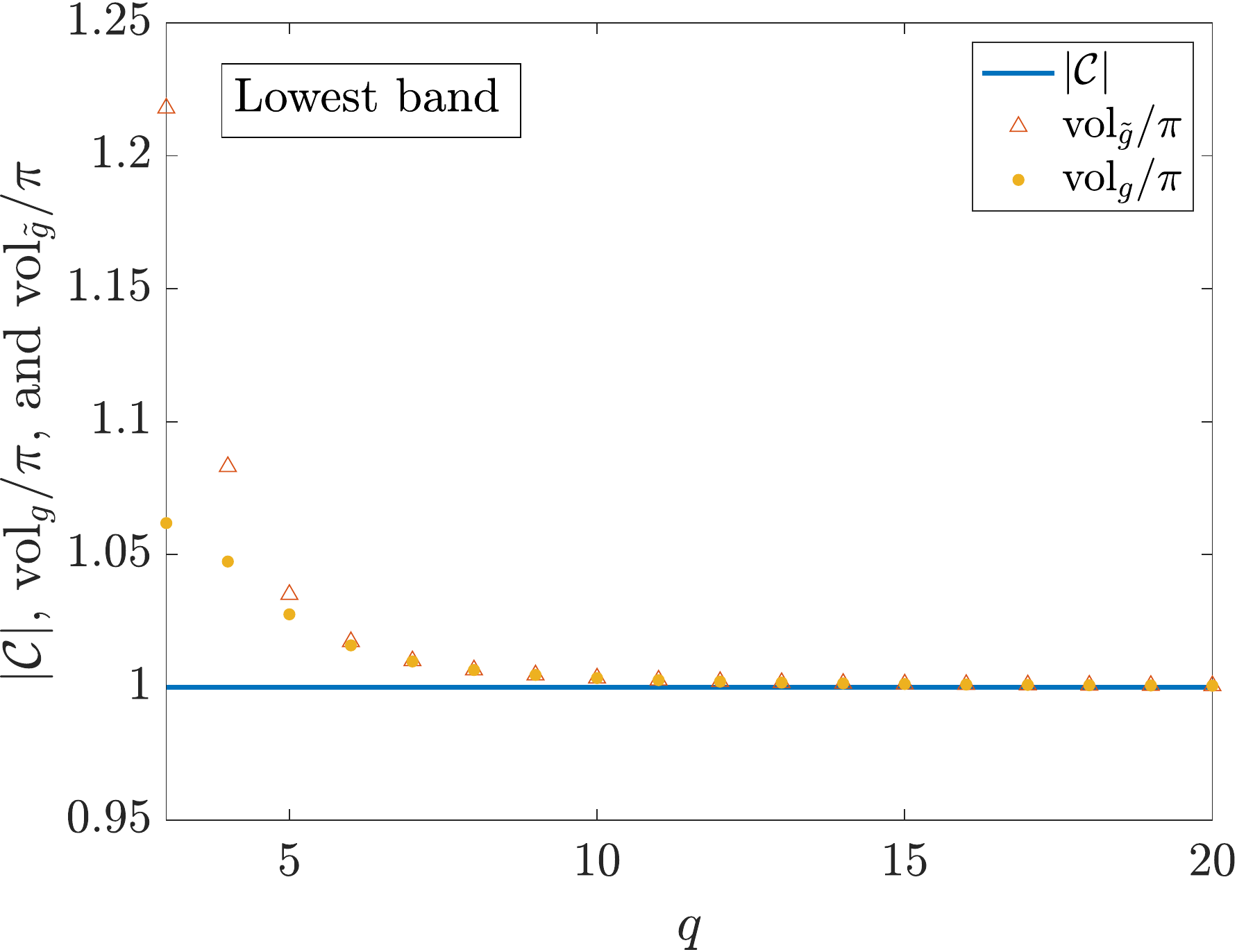}
\includegraphics[width= 0.4 \textwidth]{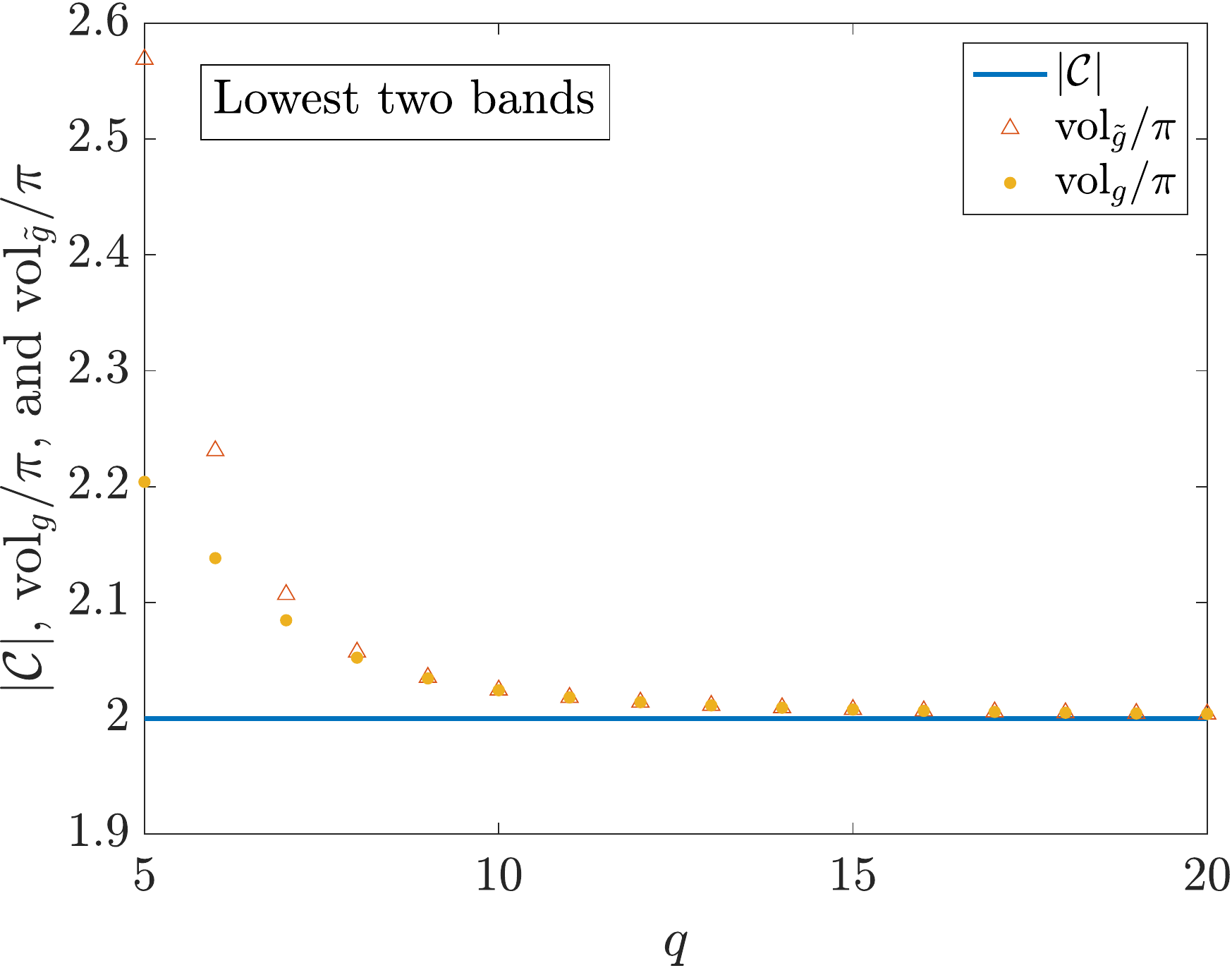}
\caption{The Harper-Hofstadter model: The Chern number and quantum volumes as a function of $q$, where $2\pi/q$ is the magnitude of the magnetic flux through a plaquette. The left figure is for the lowest band, and the right figure is for the lowest two bands. The solid line corresponds to the Chern number, and the circles and triangles are $\mathrm{vol}_g/\pi$ and $\mathrm{vol}_{\tilde{g}}/\pi$, respectively.}
\label{fig:hh}
\end{center}
\end{figure*}

First, we consider the lowest band of $\phi = 2\pi/q$. It is known in this case that the lowest band has the Chern number of $-1$. We calculate the Chern number and the quantum volume of the lowest band for various values of $q$. The results are plotted in Fig.~\ref{fig:hh}. We see that the inequality Eq.~(\ref{eq:main}) becomes saturated as one increases $q$. In the Harper-Hofstadter model, the large $q$ limit corresponds to the Landau level limit~\cite{fen:sim:roy:14}, and thus the lowest band should approach the lowest Landau level. The saturation of the inequality is consistent with the property of the lowest Landau level that the equality is achieved.

Next we consider the lowest two bands of $\phi = 2\pi/q$. When $q \ge 5$, lowest two bands both have the Chern number of $-1$, and thus the Chern number of the two bands combined is $-2$. In Fig.~\ref{fig:hh} we also plot the Chern number and the quantum volume of the lowest two bands. The inequality Eq.~(\ref{eq:main}) is saturated again in the large $q$ limit, confirming that the limit of Landau levels is achieved also when more than one band are considered.

\subsubsection{Three-band model with unit Chern number}
\begin{figure}[htbp]
\begin{center}
\includegraphics[width=0.45 \textwidth]{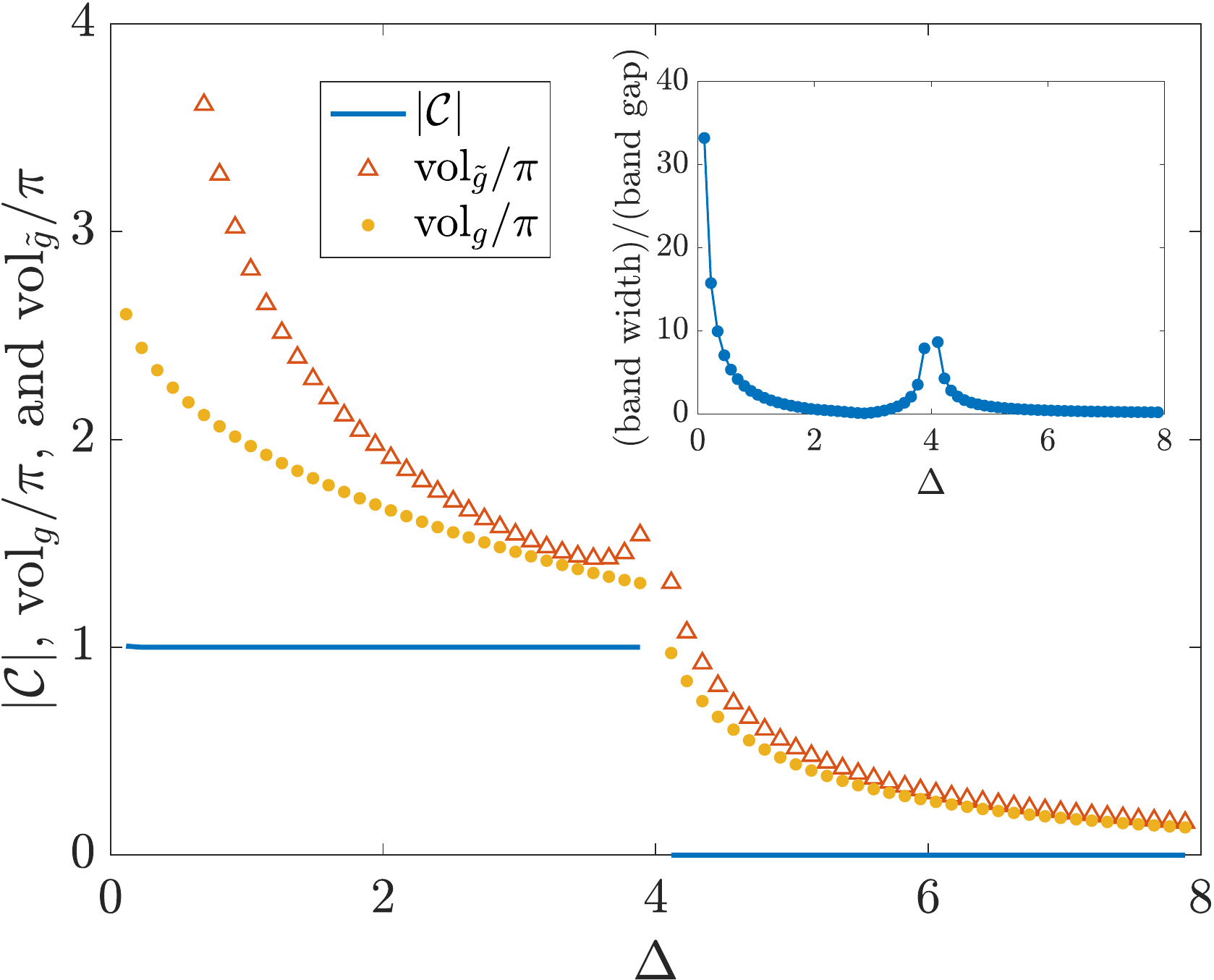}
\caption{Three band model from Sun {\it et al}.~\cite{Sun:2011}: The Chern number and quantum volumes as a function of a parameter $\Delta$. The solid line corresponds to the Chern number, and the circles and triangles are $\mathrm{vol}_g/\pi$ and $\mathrm{vol}_{\tilde{g}}/\pi$, respectively. The inset shows the band flatness, which is defined by the ratio between the bandwidth and the band gap.}
\label{fig:sun3band}
\end{center}
\end{figure}

Next, we consider the three-band model introduced in Sun {\it et al}. Ref.~\cite{Sun:2011}. In this model, one can make a band very flat by properly choosing the parameters. We investigate the relation between the band flatness and the saturation of the inequality Eq.~(\ref{eq:main}). The momentum-space Hamiltonian of the model is
\begin{widetext}
\begin{align}
	H(\mathbf{k})
	=
	\begin{pmatrix}
	-2t_{dd} (\cos k_x + \cos k_y) + \delta & 2it_{pd} \sin k_x & 2it_{pd} \sin k_y \\
	-2i t_{pd} \sin k_x & 2t_{pp} \cos k_x - 2t_{pp}^\prime \cos k_y & i\Delta \\
	-2it_{pd} \sin k_y & -i\Delta & 2t_{pp}\cos k_y - 2t^\prime_{pp} \cos k_x
	\end{pmatrix}.
\end{align}
\end{widetext}
We focus on the lowest band.
A topological phase transition can be induced by changing $\Delta$.
We use a set of parameters discussed in Ref.~\cite{Sun:2011}, which are $t_{dd} = t_{pd} = t_{pp} = 1$ and $\delta = -4t_{dd} + 2t_{pp} + \Delta - 2t_{pp} \Delta / (4 t_{pp} + \Delta)$ and $t_{pp}^\prime = t_{pp} \Delta / (4t_{pp} + \Delta)$.
The Chern number and the quantum volumes are plotted in Fig.~\ref{fig:sun3band}. There is a topological phase transition at $\Delta = 4$. We also plot the band flatness in the inset of Fig.~\ref{fig:sun3band}. The band flatness is evaluated by the ratio between the band width and the band gap; the smaller value of this ratio implies that the band is flatter. In the topologically trivial regime, the quantum volume in momentum space predicts the Chern number being zero correctly apart from region in the vicinity of the topological phase transition. In the topologically nontrivial regime, the agreement between the Chern number and the quantum volume becomes better as the band becomes flatter. We expect this tendency that flattening of a band makes the inequality Eq.~(\ref{eq:main}) saturated to hold in general from an analogy with the Landau levels, which are flat and topological.

\subsubsection{Three-band model with higher Chern numbers}
\begin{figure}[htbp]
\begin{center}
\includegraphics[width=0.45 \textwidth]{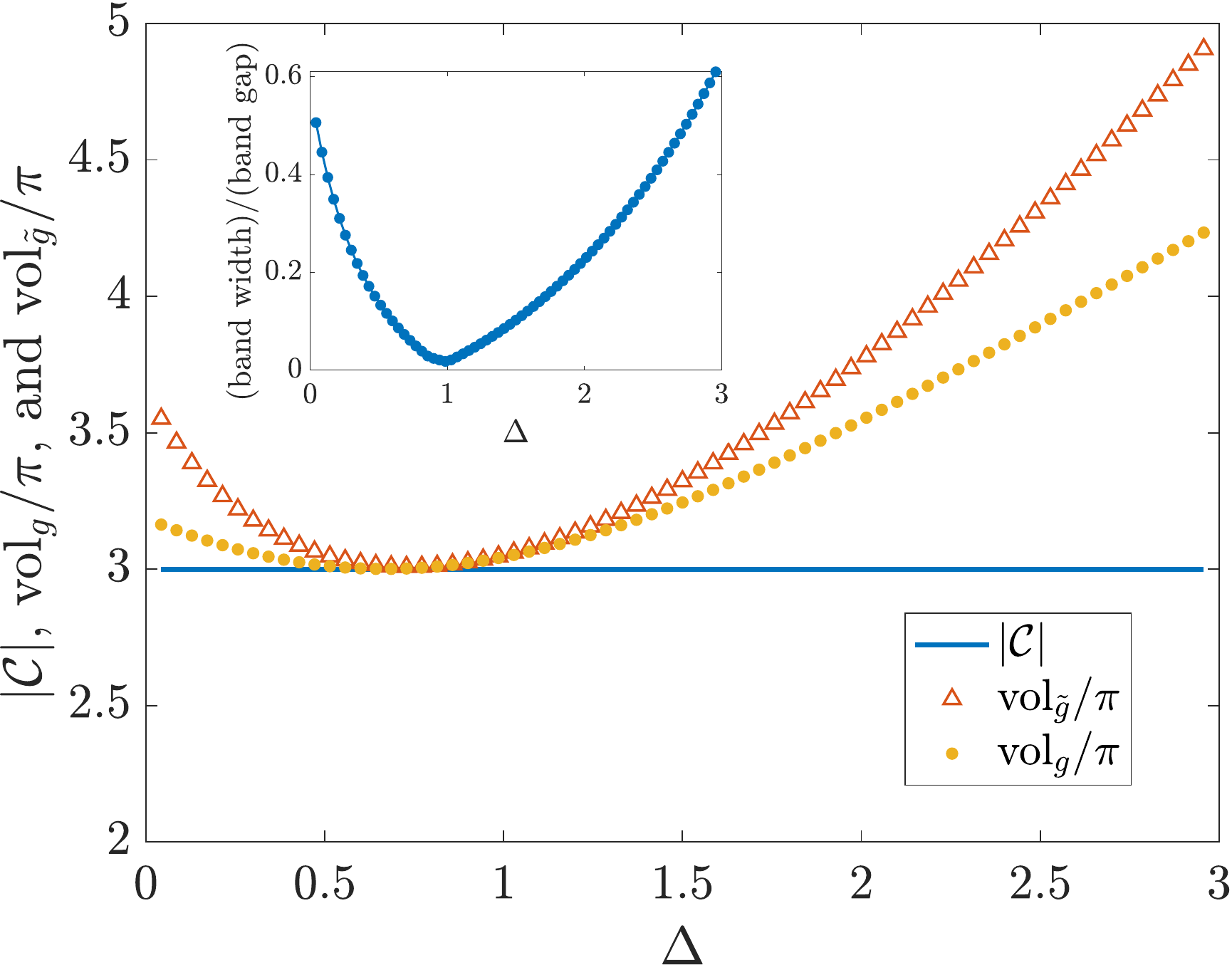}
\caption{Three-band model from Yang {\it et al}.~\cite{Yang:2012}: The Chern number and quantum volumes as a function of a parameter $\Delta$. The solid line corresponds to the Chern number, and the circles and triangles are $\mathrm{vol}_g/\pi$ and $\mathrm{vol}_{\tilde{g}}/\pi$, respectively. The inset shows the band flatness, which is defined by the ratio between the bandwidth and the bandgap.}
\label{fig:yang3band}
\end{center}
\end{figure}

The tendency that flatter bands lead to $|\mathcal{C}| \approx \mathrm{vol}_g$ also holds when we have a band with the Chern number greater than one.
To see this property, we consider another three band model proposed in Yang {\it et al}.~\cite{Yang:2012}.
The lowest band of the model has the Chern number of three, and depending on parameters, the band can become very flat. The Hamiltonian is
\begin{align}
    H(\mathbf{k})
    =
    \begin{pmatrix}
    \mathcal{F}_\mathbf{k}(\phi)/2 & \mathcal{G}_{2,\mathbf{k}}(5\phi)
    & \mathcal{G}_{1,\mathbf{k}}(6\phi) \\
    \mathcal{G}_{1,\mathbf{k}}(2\phi)& \mathcal{F}_{\mathbf{k}}(3\phi)/2
    &\mathcal{G}_{2,\mathbf{k}}(7\phi) \\
    \mathcal{G}_{2,\mathbf{k}}(3\phi) & \mathcal{G}_{1,\mathbf{k}}(4\phi)& \mathcal{F}_{\mathbf{k}}(5\phi)/2
    \end{pmatrix} + \mathrm{H.c.},
\end{align}
where
\begin{align}
    \mathcal{F}_{\mathbf{k}}(\Phi) &\equiv 2t_2 \cos (k_x + k_y - \Phi) \\
    \mathcal{G}_{1,\mathbf{k}}(\Phi) &\equiv t_1 e^{ik_x} + t_1 e^{-ik_y + i\Phi} \\
    \mathcal{G}_{2,\mathbf{k}}(\Phi) &\equiv t_2 e^{ik_x - ik_y + i\Phi}.
\end{align}
As in Ref.~\cite{Yang:2012}, we take $t_1 = 1$ and $\phi = \pi/3$. We take $t_2 = -\Delta/\sqrt{3}$, and vary $\Delta$ to tune the band flatness of the lowest band.
In Fig.~\ref{fig:yang3band}, we plot the Chern number and quantum volumes as a function of the tuning parameter $\Delta$, as well as the band flatness. The band becomes most flat when $\Delta = 1$, and indeed the equality $|\mathcal{C}| \approx \mathrm{vol}_g/\pi \approx \mathrm{vol}_{\tilde{g}}/\pi$ almost holds around the region where the band is flat. As the band flatness is reduced, the discrepancy among $|\mathcal{C}|$, $\mathrm{vol}_g/\pi$, and $\mathrm{vol}_{\tilde{g}}/\pi$ becomes larger.
With these examples, we can heuristically say that the inequality Eq.~(\ref{eq:main}) approaches equality when the band becomes flatter.

\section{Conclusion}
\label{sec:conclusion}
Although Chern number and the quantum metric are both derived from the same map from the parameter space to the space of quantum states, their physical manifestations are quite different. As a bulk response, the former is related to an anti-symmetric responses such as the Hall conductivity and anomalous velocity, whereas the latter is more related to the localization properties of the system. Despite the difference, we have seen that because of the inequality
\begin{align}
    \pi |\mathcal{C}| \le \mathrm{vol}_g \le \mathrm{vol}_{\tilde{g}} \label{eq:last}
\end{align}
we can infer topological properties just by looking at the quantum metric and vice versa.
Because of the inequality, the quantum volume of momentum space, $\mathrm{vol}_g$, is always closer to $\pi |\mathcal{C}|$ than the quantum volume of the twist-angle space, $\mathrm{vol}_{\tilde{g}}$. On the other hand, we note that when one has a fermionic band insulator, what is experimentally relevant is often the quantum metric in the twist-angle space, $\mathrm{vol}_{\tilde{g}}$, as manifested in the experiment Ref.~\cite{Asteria:2019}. Since each term of the inequality Eq.~(\ref{eq:last}) is related to observables in different experimental setups, the inequality provides us the possibility of studying topology and geometry from multiple perspectives.

If similar inequalities hold for higher-dimensional Chern insulators and/or symmetry-protected topological phases is left for future studies. Furthermore, as discussed in more detail in the accompanying paper~\cite{MeraOzawa:published}, the saturation of the inequality is related to holomorphic structure of the map from the parameter space to the space of the quantum states. Studying consequences of the inequality both from physical and mathematical points of view will open a new avenue towards a unified understanding of the two apparently different concepts of topology and localization.

\begin{acknowledgments}
We thank Hosho Katsura for suggesting a simple proof for Appendix A.
T.O. acknowledges support from JSPS KAKENHI Grant No. JP20H01845, JST PRESTO Grant No. JPMJPR19L2, JST CREST Grant No. JPMJCR19T1, and RIKEN iTHEMS. B.M. acknowledges the support from SQIG -- Security and Quantum Information Group, the Instituto de Telecomunica\c{c}\~oes (IT) Research Unit, Ref. UIDB/50008/2020, funded by Funda\c{c}\~ao para a Ci\^{e}ncia e a Tecnologia (FCT), European funds, namely, H2020 project SPARTA, as well as  projects QuantMining POCI-01-0145-FEDER-031826 and PREDICT PTDC/CCI-CIF/29877/2017.
\end{acknowledgments}

\appendix
\section{Proof of $\sqrt{\det (g)} \ge |\Omega_{12}|/2$}
\label{sec:app1}
In this appendix, we give a proof of the inequality Eq.(\ref{eq:ineq}), i.e., $\sqrt{\det (g(\boldsymbol\lambda))} \ge |\Omega_{12}(\boldsymbol\lambda)|/2$ identifying the condition under which the inequality is saturated.
Let us consider two sets of vectors
\begin{align}
    |\alpha\rangle
    &=
    \begin{pmatrix}
    |\alpha_1\rangle \\
    |\alpha_2\rangle \\
    \vdots \\
    |\alpha_r\rangle
    \end{pmatrix},
    &
    |\beta\rangle
    &=
    \begin{pmatrix}
    |\beta_1\rangle \\
    |\beta_2\rangle \\
    \vdots \\
    |\beta_r\rangle
    \end{pmatrix},
\end{align}
where each $|\alpha_m\rangle$ and $|\beta_m\rangle$ is an element of the Hilbert space spanned by the quantum states for a given value of $\boldsymbol\lambda$. As in the main text, $r$ denotes the number of occupied bands.
Let us define a complex inner product $\langle \alpha |\beta\rangle$ between two sets of vectors by
\begin{align}
	\langle \alpha|\beta\rangle \equiv \sum_{m = 1}^r \langle \alpha_m| \beta_m \rangle.
\end{align}
It is straightforward to show that this definition fulfills the condition of inner product.
Then, the following Cauchy-Schwarz inequality holds~\cite{PrincetonCompanion}:
\begin{align}
    \langle \alpha | \alpha \rangle \langle \beta | \beta \rangle \ge |\langle \alpha | \beta \rangle|^2, \label{eq:cs}
\end{align}
and the equality holds if either $|\alpha\rangle$ or $|\beta\rangle$ is zero, or if there exists a complex number $c$, which satisfies $|\alpha\rangle = c |\beta \rangle$.
We now take
\begin{align}
	|\alpha_m \rangle &= (1-P) |\partial_i \psi_m (\boldsymbol\lambda)\rangle, \\
	|\beta_m \rangle &= (1-P) |\partial_j \psi_m (\boldsymbol\lambda) \rangle,
\end{align}
where $P$ is, as in the main text, the projector $P = \sum_{m=1}^r |\psi_m(\boldsymbol\lambda)\rangle\langle \psi_m (\boldsymbol\lambda)|$.
Using the property $(1-P)^2 = (1-P)$, one can see that
\begin{align}
    \langle \alpha | \alpha\rangle &= \chi_{ii} (\boldsymbol\lambda),
    &
    \langle \beta | \beta\rangle &= \chi_{jj} (\boldsymbol\lambda),
    &
    \langle \alpha | \beta\rangle &= \chi_{ij} (\boldsymbol\lambda).
\end{align}
The Cauchy-Schwarz inequality Eq.~(\ref{eq:cs}) then gives us
\begin{align}
	\chi_{ii}(\boldsymbol\lambda) \chi_{jj}(\boldsymbol\lambda) \ge |\chi_{ij}(\boldsymbol\lambda)|^2.
\end{align}
Note that, for two-dimensional parameter spaces, this relation is nothing but the positive semi-definiteness of the quantum geometric tensor $\det (\chi) \ge 0$~\cite{Peotta:2015}.
Since $\chi_{ii} = g_{ii}$ and $|\chi_{ij}(\boldsymbol\lambda)|^2 = (\mathrm{Re}[\chi_{ij}])^2 + (\mathrm{Im}[\chi_{ij}])^2 = g_{ij}^2 + \Omega_{ij}^2/4$, we obtain
\begin{align}
	g_{ii}g_{jj} - g_{ij}^2 \ge \Omega_{ij}^2/4
\end{align}
Now, considering a two-dimensional parameter space and taking $i = 1$ and $j = 2$, we obtain
\begin{align}
	\det (g) \ge |\Omega_{12}|^2/4.
\end{align}
Taking the square root of both sides, we obtain the desired inequality.
The condition for the inequality to be saturated is either $\chi_{11} = \chi_{12} = \chi_{22} = 0$ or
$\exists\, c \in \mathbb{C}$ such that $\forall m,\ (1 - P)|\partial_1 \psi_m (\boldsymbol\lambda)\rangle = c (1-P)|\partial_2 \psi_m (\boldsymbol\lambda) \rangle$. One can see that, for two-band models, the inequality is always saturated because $1-P$ projects to a one-dimensional subspace, and thus the condition $\exists\, c \in \mathbb{C}$ such that $(1 - P)|\partial_1 \psi_1 (\boldsymbol\lambda)\rangle = c (1-P)|\partial_2 \psi_1 (\boldsymbol\lambda) \rangle$ always holds regardless of the details of $|\psi_1 (\boldsymbol\lambda)\rangle$.

We also note that, since the two eigenvalues of the matrix $g$ are nonnegative, $\mathrm{tr}(g) \ge 2\sqrt{\det (g)}$ holds in general. This implies $\mathrm{tr}(g) \ge 2 \sqrt{\det (g)} \ge |\Omega_{12}|$, which was also first noted by Roy~\cite{Roy:2014}.

\section{Proof of $\mathrm{vol}_g \le \mathrm{vol}_{\tilde{g}}$}
\label{sec:gg}

We now prove $\mathrm{vol}_g \le \mathrm{vol}_{\tilde{g}}$, which states that the quantum volume of momentum space is smaller or equal to the quantum volume of the twist-angle space. Since the quantum metric in the twist-angle space,  $\tilde{g}_{ij}(\theta_x, \theta_y)$, is flat, we have $\mathrm{vol}_{\tilde{g}} = (2\pi)^2 \det (\tilde{g}_{ij})$. Then, explicitly writing out the expressions for the determinant, the statement we want to prove can be rewritten in the following form:
\begin{align}
	&\mathrm{vol}_g = \int d^2 k \sqrt{g_{xx}g_{yy} - g_{xy}^2}
	\notag \\
	&\le \sqrt{\int d^2 k g_{xx} \int d^2 k g_{yy} - \left( \int d^2 k g_{xy} \right)^2} = \mathrm{vol}_{\tilde{g}}.
\end{align}
To prove this relation, we use the Cauchy-Schwarz inequality again, which, in a formulation relevant to the current case, states that for any square-integrable real nonnegative functions $\phi(\mathbf{k})$ and $\psi (\mathbf{k})$,
\begin{align}
	\int d^2 k \; \phi(\mathbf{k}) \psi(\mathbf{k}) \le \sqrt{\int d^2 k [\phi (\mathbf{k})]^2 \cdot \int d^2 k [\psi (\mathbf{k})]^2}.
\end{align}
To use the Cauchy-Schwarz inequality, we write
\begin{align}
	\mathrm{vol}_g = \int d^2 k \sqrt{\sqrt{g_{xx}g_{yy}} - g_{xy}} \cdot \sqrt{\sqrt{g_{xx}g_{yy}} + g_{xy}}.
\end{align}
Using the Cauchy-Schwarz inequality, we obtain
\begin{align}
	\mathrm{vol}_g &\le \sqrt{\int d^2k \left( \sqrt{g_{xx}g_{yy}} - g_{xy} \right) \cdot \int d^2k \left( \sqrt{g_{xx}g_{yy}} + g_{xy} \right)}
	\notag \\
	&=
	\sqrt{\left( \int d^2k \sqrt{g_{xx}g_{yy}}  \right)^2 - \left( \int d^2k g_{xy} \right)^2}.
\end{align}
Using again the Cauchy-Schwarz inequality to the first term inside the square root, we obtain the desired relation:
\begin{align}
	\mathrm{vol}_g &\le
	\sqrt{\int d^2 k g_{xx} \int d^2k g_{yy} - \left( \int d^2k g_{xy} \right)^2}
	=
	\mathrm{vol}_{\tilde{g}}.
\end{align}
The condition for the Cauchy-Schwarz inequality to hold as an equality, when both functions $\phi$ and $\psi$ are not identically zero, is that the two functions are linearly dependent, namely, one is a scalar multiple of the other.
Upon deriving $\mathrm{vol}_g\le \mathrm{vol}_{\tilde{g}}$, we used the Cauchy-Schwarz inequality twice. For the equality to hold in both steps, we need that $g_{xy}$ and $g_{yy}$ are both scalar multiples of $g_{xx}$. This is equivalent to saying that the quantum metric in momentum space takes the following form:
\begin{align}
	g (k_x, k_y)
	&= f (k_x, k_y) \begin{pmatrix} c_{xx} & c_{xy} \\ c_{xy} & c_{yy} \end{pmatrix},
\end{align}
where $c_{xx}$, $c_{xy}$, and $c_{yy}$ are $\mathbf{k}$-independent real numbers and $f (k_x, k_y)$ is a nonnegative function.

\section{Geometry and topology of Landau levels}
\label{sec:ll}

Here we briefly summarize the derivation of the geometrical properties and the first Chern number of Landau levels.
For single Landau levels, the similar quantities have also been evaluated in Ref.~\cite{Peotta:2015} by taking appropriate limits of the Harper-Hofstadter model. In the alternative derivation given here, we work directly on the Landau level wave functions, and provide the quantum geometric tensor of a collection of $r$ Landau levels.
The Hamiltonian of a charged particle in a uniform magnetic field $B$ in a two-dimensional space is given by
\begin{align}
	H = \frac{(\hat{p}_x - \hat{A}_x)^2 + (\hat{p}_y - \hat{A}_y)^2}{2M},
\end{align}
with a magnetic vector potential $\mathbf{A} = (\hat{A}_x, \hat{A}_y)$ satisfying $B = \partial_x \hat{A}_y - \partial_y \hat{A}_x$. Later we take the Landau gauge $\mathbf{A} = (0, B\hat{x})$. We take the charge to be unity.
This Hamiltonian has a translational symmetry only in $y$ direction, and the momentum along $x$ direction is not a good quantum number. We can, however, consider a magnetic translational symmetry, which is a translation followed by a phase shift. Using the magnetic translational symmetry, we can label eigenstates using quasimomenta along both $x$ and $y$ directions, and the geometrical and topological properties can be defined as in Sec.~\ref{sec:defs}.

To introduce the magnetic translational symmetry, we consider a fictitious magnetic unit cell, which is a rectangle of length $a_x$ in $x$ direction and $a_y$ in $y$ direction satisfying
\begin{align}
    |B| a_x a_y = 2\pi. \label{eq:magunit}
\end{align}
Note that the choice of $a_x$ and $a_y$ satisfying the relation Eq.~(\ref{eq:magunit}) is not unique, and the result below does not depend on the choice.

A lattice vector $\mathbf{R}$ is any linear combination of $\mathbf{a}_x \equiv (a_x,0)$ and $\mathbf{a}_y \equiv (0,a_y)$ with integer coefficients.
Now, we define a usual translation operator by
\begin{align}
	\mathcal{T}(\mathbf{R}) \equiv e^{i\mathbf{R}\cdot\hat{\mathbf{p}}} = e^{\mathbf{R}\cdot\nabla},
\end{align}
which satisfies
\begin{align}
	\mathcal{T}(\mathbf{R}) f(\mathbf{r})
	=
	f(\mathbf{r} + \mathbf{R}),
\end{align}
when acted to a smooth function $f(\mathbf{r})$. The translational operator $\mathcal{T}(\mathbf{R})$ does not commute with the Hamiltonian.
However, by introducing a function $\xi_\mathbf{R}(\mathbf{r})$, which satisfies
\begin{align}
	\mathbf{A}(\mathbf{r} + \mathbf{R})
	=
	\mathbf{A}(\mathbf{r}) + \nabla \xi_\mathbf{R} (\mathbf{r}), \label{xidef}
\end{align}
we can define a magnetic translation operator:
\begin{align}
	\mathcal{M}(\mathbf{R})
	\equiv
	e^{-i\xi_\mathbf{R}(\mathbf{r})}\mathcal{T}(\mathbf{R}). \label{magtdef}
\end{align}
The magnetic translation operators $\mathcal{M}(\mathbf{R})$ commute with themselves and also with the Hamiltonian. In addition, one can show $\mathcal{M}(\mathbf{R}_1)\mathcal{M}(\mathbf{R}_2) = \mathcal{M}(\mathbf{R}_1 + \mathbf{R}_2)$.
With these properties, we can prove the Bloch theorem for the magnetic translation operators.

The Bloch theorem now states that the Hamiltonian and the magnetic translation operators can be simultaneously diagonalized. An eigenstate $\psi_{m,\mathbf{k}} (\mathbf{r})$ is characterized by a band index $m$ and a quasimomentum $\mathbf{k}$ and satisfies
\begin{align}
	\mathcal{M}(\mathbf{R}) \psi_{m,\mathbf{k}} (\mathbf{r})
	=
	e^{i\mathbf{k}\cdot \mathbf{R}} \psi_{m,\mathbf{k}} (\mathbf{r}).
\end{align}
On the other hand, using (\ref{magtdef}), one obtains
\begin{align}
	\mathcal{M}(\mathbf{R}) \psi_{m,\mathbf{k}} (\mathbf{r})
	&=
	e^{-i\xi_\mathbf{R}(\mathbf{r})}\mathcal{T}(\mathbf{R}) \psi_{m,\mathbf{k}} (\mathbf{r})
	\notag \\
	&=
	e^{-i\xi_\mathbf{R}(\mathbf{r})}\psi_{m,\mathbf{k}} (\mathbf{r} + \mathbf{R}).
\end{align}
This means
\begin{align}
	\psi_{m,\mathbf{k}} (\mathbf{r} + \mathbf{R})
	=
	e^{i\mathbf{k}\cdot \mathbf{R} + i\xi_\mathbf{R}(\mathbf{r})} \psi_{m,\mathbf{k}} (\mathbf{r}). \label{cond1}
\end{align}
The Bloch wave function $u_{m,\mathbf{k}}(\mathbf{r})$ is defined by
\begin{align}
	\psi_{m,\mathbf{k}} (\mathbf{r})
	=
	e^{i\mathbf{k}\cdot \mathbf{r}}u_{m,\mathbf{k}}(\mathbf{r}). \label{cond2}
\end{align}
From (\ref{cond1}) and (\ref{cond2}), one obtains the twisted boundary condition for the Bloch wave functions upon translation by a lattice vector:
\begin{align}
	u_{m,\mathbf{k}}(\mathbf{r} + \mathbf{R})
	=
	e^{i\xi_\mathbf{R}(\mathbf{r})}u_{m,\mathbf{k}}(\mathbf{r}). \label{eq:bc}
\end{align}
In the absence of the magnetic field, the Bloch wave function satisfies $u_{m,\mathbf{k}}(\mathbf{r} + \mathbf{R}) = u_{m,\mathbf{k}}(\mathbf{r})$. So, the additional factor of $e^{i\xi_\mathbf{R}(\mathbf{r})}$ is the peculiarity of having nonzero magnetic field.
We note that the condition Eq.~(\ref{eq:bc}) defines the sections of a non-trivial line bundle over the unit magnetic cell in real space, which topologically is a torus. The quantities $\{e^{i\xi_{\bf{R}}}\}$, where $\bf{R}$ ranges over the lattice, form what is known in the mathematics literature as a system of multipliers for the line bundle.

Now, let us explicitly take the Landau gauge $\mathbf{A}(\mathbf{r}) = (0,Bx)$. Then we can take $\xi_{\mathbf{a}_x}(\mathbf{r}) = Ba_x y$ and $\xi_{\mathbf{a}_y}(\mathbf{r}) = 0$.
The boundary condition for the Bloch wave function for the Landau gauge is then
\begin{align}
	u_{m,\mathbf{k}}(x + a_x,y) &= e^{iBa_x y}u_{m,\mathbf{k}}(x,y), \notag \\
	u_{m,\mathbf{k}}(x,y + a_y) &= u_{m,\mathbf{k}}(x,y). \label{bcon}
\end{align}

We now want to find eigenstates of $H$, which satisfy
\begin{align}
	H e^{i\mathbf{k}\cdot \mathbf{r}}u_{m,\mathbf{k}}(\mathbf{r})
	=
	E_{m,\mathbf{k}} e^{i\mathbf{k}\cdot \mathbf{r}}u_{m,\mathbf{k}}(\mathbf{r}) \label{getit}
\end{align}
and the boundary conditions (\ref{bcon}). Once we find the Bloch wave functions $u_{m,\mathbf{k}}(\mathbf{r})$, it is straightforward to calculate the Berry curvature and the Chern number from their definitions.

Under the Landau gauge, the momentum along $y$ direction is a good quantum number.
Writing an eigenstate as $e^{ik_y y} \psi(x,t)$, the Schr\"odinger equation becomes
\begin{align}
	i\frac{\partial}{\partial t} e^{ik_y y} \psi(x,t)
	=
	\frac{\hat{p}_x^2 + (k_y - B\hat{x})^2}{2M}e^{ik_y y} \psi(x,t),
\end{align}
This equation can be rewritten to become
\begin{align}
	i\frac{\partial}{\partial t} \psi(x,t)
	=
	\left[\frac{\hat{p}_x^2}{2M} +  \frac{B^2}{2M}\left(\hat{x} - \frac{k_y}{B}\right)^2 \right] \psi(x,t).
\end{align}
This is the Schr\"odinger equation of a particle in a one-dimensional harmonic oscillator with the frequency $\omega = |B|/M$ and the origin shifted at $x_0 = k_y/B$. This means that, for a given value of $k_y$, the energy level is $\omega (m+1/2)$ with an integer $m \ge 0$.
We note that the lowest Landau level is given by $m = 0$.
The (unnormalized) eigenstate is
\begin{align}
	h_m (x - k_y/B)
	=
	e^{-(x-k_y/B)^2/2l_B^2} \mathrm{H}_m ((x-k_y/B)/l_B), \label{eigenstate}
\end{align}
where $l_B \equiv 1/\sqrt{|B|}$ is the magnetic length and $H_m$ is the Hermite polynomial with degree $m$.
This eigenstate does not obey the periodicity of the Bloch wave function Eq.~(\ref{bcon}). To obtain the Bloch wave function, we consider the following linear combination of states where values of $k_y$ are separated by $2\pi/a_y$ times an integer:
\begin{align}
	&\psi_{m,\mathbf{k}}(\mathbf{r})
	\notag \\
	&\equiv
	\sum_{l=-\infty}^\infty e^{ik_x a_x l} e^{i(k_y + 2\pi l/a_y) y} h_m (x-(k_y+ 2\pi l/a_y)/B)
	\notag \\
	&= \sum_{l=-\infty}^\infty e^{ik_x a_x l} e^{i(k_y + 2\pi l/a_y) y} h_m (x- a_x l - k_y/B)
\end{align}
Each term in the sum is an eingenstate of the original Hamiltonian with the eigenvalue $\omega(m+1/2)$, so the linear combiation $\psi_{m,\mathbf{k}}(\mathbf{r})$ is still an eigenstate of the original Hamiltonian with the same eigenvalue:
\begin{align}
	H \psi_{m,\mathbf{k}}(\mathbf{r}) = \omega \left( m + \frac{1}{2} \right) \psi_{m,\mathbf{k}}(\mathbf{r}).
\end{align}
We now define
\begin{align}
	u_{m,\mathbf{k}}(\mathbf{r})
	&\equiv
	e^{-i\mathbf{k}\cdot \mathbf{r}}\psi_{m,\mathbf{k}}(\mathbf{r}) \label{correctbloch}
	\\
	&=
	\sum_{l=-\infty}^\infty e^{-ik_x (x - a_x l) + i 2\pi l y/a_y} h_m (x- a_x l - k_y/B).
	\notag
\end{align}
One can confirm that this function $u_{m,\mathbf{k}}(\mathbf{r})$ is the desired Bloch wave function, which obeys the boundary condition Eq.~(\ref{bcon}).

Using the Bloch wave function (\ref{correctbloch}), we can now calculate the quantum geometric tensor and the Chern number.
We note that the Bloch wave function (\ref{correctbloch}) is not normalized.
We introduce an unnormalized Bloch wave function in a ``ket" form $|\tilde{u}_{m,\mathbf{k}}\rangle$ which satisfies, in position representation, $\langle \mathbf{r}|\tilde{u}_{m,\mathbf{k}}\rangle = u_{m,\mathbf{k}(\mathbf{r})}$.
Its inner product with itself satisfies
\begin{align}
    \langle \tilde{u}_{m,\mathbf{k}}|\tilde{u}_{m,\mathbf{k}}\rangle
    =
    \int_{0}^{a_x}dx \int_{0}^{a_y}dy |u_{m,\mathbf{k}}(x,y)|^2.
\end{align}
In terms of the unnormalized Bloch wave function, one can show that the quantum geometric tensor when the lowest $r$ bands are occupied takes the following form:
\begin{align}
    \chi_{ij}(\mathbf{k}) = \sum_{m = 0}^{r-1}
    \frac{\langle\partial_{k_i} \tilde{u}_{m,\mathbf{k}}| ( 1 - P(\mathbf{k}))|\partial_{k_j}\tilde{u}_{m,\mathbf{k}}\rangle}{\langle \tilde{u}_{m,\mathbf{k}}|\tilde{u}_{m,\mathbf{k}}\rangle}. \label{eq:chiun}
\end{align}
where
\begin{align}
    P(\mathbf{k}) = \sum_{m=0}^{r-1}\frac{|\tilde{u}_{m,\mathbf{k}}\rangle \langle \tilde{u}_{m,\mathbf{k}}|}{\langle \tilde{u}_{m,\mathbf{k}}|\tilde{u}_{m,\mathbf{k}}\rangle}.
\end{align}
The denominators take care of the normalization of the wave function.

The matrix elements can be calculated using standard properties of Hermite polynomials. For example,
\begin{align}
    \langle \tilde{u}_{m,\mathbf{k}}|\tilde{u}_{m,\mathbf{k}}\rangle
    &= \int_{0}^{a_x}dx \int_{0}^{a_y}dy |u_{m,\mathbf{k}}(x,y)|^2
    \notag \\
    &=
    a_y \int_0^{a_x} dx \sum_{l = -\infty}^{\infty} h_m (x - a_x l - k_y/B)^2
    \notag \\
    &=
    a_y \int_{-\infty}^\infty h_m (x - k_y/B)^2
    =
    a_y \int_{-\infty}^\infty h_m (x)^2
    \notag \\
    &=
    a_y l_B \int_{-\infty}^\infty dX d^{-X^2}H_n(X)^2 \notag \\
    &=
    a_y l_B \sqrt{\pi} 2^n n!. \label{eq:den}
\end{align}
Also,
\begin{align}
    &\langle \partial_{k_x} \tilde{u}_{m,\mathbf{k}} |\partial_{k_x} \tilde{u}_{m,\mathbf{k}}\rangle
    \notag \\
    &=
    \int_0^{a_x} dx \int_0^{a_y}dy
    \partial_{k_x} u^*_{m,\mathbf{k}}(x,y)
    \partial_{k_x} u_{m,\mathbf{k}}(x,y)
    \notag \\
    &=
    a_y\int_0^{a_x} dx \sum_{l = -\infty}^{\infty} (x - a_xl)^2 h_m (x - a_x l -k_y/B)^2
    \notag \\
    &=
    a_y \int_{-\infty}^{\infty}dx (x + k_y/B)^2 h_m (x)^2
    \notag \\
    &=
    a_y \int_{-\infty}^{\infty}dx (x^2 + k_y^2/B^2) h_m (x)^2,
\end{align}
where in the final step we used that $h_m (x)^2$ is an even function of $x$.
Using the recurrence relations of the Hermite polynomial, one can show
\begin{align}
    &x^2 h_m (x)
    \\
    &= l_B^2 \left[ \frac{1}{4}h_{m+2}(x) + \left( m + \frac{1}{2} \right) h_m (x) + m(m-1) h_{m-2}(x) \right].\notag
\end{align}
Then, using the orthogonality of the Hermite polynomials, we obtain
\begin{align}
     \frac{\langle \partial_{k_x} \tilde{u}_{m,\mathbf{k}} |\partial_{k_x} \tilde{u}_{m,\mathbf{k}}\rangle}{\langle \tilde{u}_{m,\mathbf{k}}|\tilde{u}_{m,\mathbf{k}}\rangle}
     =
     \frac{1}{|B|} \left( m + \frac{1}{2} \right) + \frac{k_y^2}{B^2}.
\end{align}
Similarly, one can show
\begin{align}
	\frac{\langle \partial_{k_y} \tilde{u}_{m,\mathbf{k}} |\partial_{k_y} \tilde{u}_{m,\mathbf{k}}\rangle}{\langle \tilde{u}_{m,\mathbf{k}}|\tilde{u}_{m,\mathbf{k}}\rangle}
	&=
	\frac{1}{|B|}\left( m + \frac{1}{2}\right)
	\\
	\frac{\langle \partial_{k_x} \tilde{u}_{m,\mathbf{k}} |\partial_{k_y} \tilde{u}_{m,\mathbf{k}}\rangle}{\langle \tilde{u}_{m,\mathbf{k}}|\tilde{u}_{m,\mathbf{k}}\rangle}
	&=
	\frac{i}{2B}.
\end{align}
Furthermore, the inter-band matrix elements are
\begin{align}
    &\frac{\langle \tilde{u}_{m^\prime,\mathbf{k}} |\partial_{k_x} \tilde{u}_{m,\mathbf{k}}\rangle}{\langle \tilde{u}_{m,\mathbf{k}}|\tilde{u}_{m,\mathbf{k}}\rangle} \notag \\
    &=
    -i\left( l_B (m+1)\delta_{m+1,m^\prime} + (l_B/2)\delta_{m-1,m^\prime} + (k_y/B)\delta_{m,m^\prime}\right)
    \\
    &\frac{\langle \tilde{u}_{m^\prime,\mathbf{k}} |\partial_{k_y} \tilde{u}_{m,\mathbf{k}}\rangle}{\langle \tilde{u}_{m,\mathbf{k}}|\tilde{u}_{m,\mathbf{k}}\rangle} \notag \\
    &=
    -\frac{1}{l_B B}\left( -(m+1)\delta_{m+1,m^\prime} + (1/2)\delta_{m-1,m^\prime} \right).
\end{align}
From these relations, we can compute the full quantum geometric tensor.
Below we first consider the quantum geometric tensor or each band, and then consider the quantum geometric tensor of multiple bands combined.
\subsection{Single band}
We first derive the quantum geometric tensor when only single $m$-th band is in consideration.
The quantum metric of $m$-th band is
\begin{align}
	g_{m,xx}(\mathbf{k}) &= g_{m,yy}(\mathbf{k}) = \frac{1}{|B|}\left( m + \frac{1}{2}\right),
	\\
	g_{m,xy}(\mathbf{k}) &= 0,
\end{align}
which has been obtained in Ref.~\cite{Peotta:2015}.
The quantum volume for each band is then
\begin{align}
    \mathrm{vol}_g
    &=
    \int_{-\pi/a_x}^{\pi/a_x}dk_x \int_{-\pi/a_y}^{\pi/a_y}dk_y \sqrt{\det (g)}
    \notag \\
    &=
    \int_{-\pi/a_x}^{\pi/a_x}dk_x \int_{-\pi/a_y}^{\pi/a_y}dk_y \frac{1}{|B|}\left( m + \frac{1}{2}\right)
    \notag \\
    &=
    \frac{2\pi}{a_x} \frac{2\pi}{a_y}
    \frac{1}{|B|}\left( m + \frac{1}{2}\right)
    =
    2\pi \left( m + \frac{1}{2}\right).
\end{align}
Note that since the quantum metric is flat in momentum space, the quantum volume of the twist-angle space takes the same value: $\mathrm{vol}_g = \mathrm{vol}_{\tilde{g}}$.

For the Berry curvature,
\begin{align}
	\Omega_{xy}(k_x,k_y)
	=
	-\frac{1}{B}.
\end{align}
The Chern number is then
\begin{align}
	\mathcal{C}
	&=
	\frac{1}{2\pi}\int_{-\pi/a_x}^{\pi/a_x}dk_x \int_{-\pi/a_y}^{\pi/a_y}dk_y \Omega_z(k_x,k_y)
	\notag \\
	&=
	\frac{1}{2\pi}\frac{2\pi}{a_x} \frac{2\pi}{a_y}
	\frac{-1}{B}
	=
	-\mathrm{sign}(B).
\end{align}
Therefore, the Berry curvature of Landau levels is a constant $-1/B$ irrespective of the band index $m$, and the absolute value of the Chern number is always one.

\subsection{Multiple bands}
We finally consider the situation where the lowest $r$ bands are all occupied, and thus the quantum geometric tensor is given by the expression Eq.~(\ref{eq:chiun}). From a straightforward calculation, one obtains
\begin{align}
    g_{xx}(\mathbf{k}) &= g_{yy}(\mathbf{k}) = \frac{r}{2|B|},
    &
    g_{xy}(\mathbf{k}) &= 0, &
    \Omega_{xy} &= -\frac{r}{B}.
\end{align}
Therefore, the quantum volume and the Chern number are
\begin{align}
    \mathrm{vol}_g &= \mathrm{vol}_{\tilde{g}} = r\pi, &
    \mathcal{C} &= -r\ \mathrm{sign}(B).
\end{align}
We see that the equality $\mathrm{vol}_g = \mathrm{vol}_{\tilde{g}} = \pi |\mathcal{C}|$ holds for the collection of the lowest $r$ Landau levels.
The fact that the equality $\mathrm{vol}_g = \mathrm{vol}_{\tilde{g}} = \pi |\mathcal{C}|$ holds even though each single Landau level does not satisfy the equality apart form the lowest Landau level reflects the property that the quantum metric and quantum volume of combined bands is not just a sum of each band~\cite{Peotta:2015}.

\bibliography{bibliography}

\end{document}